\documentclass[12pt,a4paper]{article}
\usepackage{amsfonts}
\usepackage{epsfig}
\def \YM	{Yang-Mills }

\def \as	{\alpha_{_{\rm{s}}}}
\def \d3x 	{d^3x}

\def \Gn	{G_{\mathcal N}}
\def \Gs	{G_{\mathcal S}}
\def \Lam	{\Lambda}

\def \rg	{\sqrt{g}}
\def \rG	{\sqrt{G}}

\def \M		{\mathcal M}

\def \AAm	{A^a_\mu}
\def \AAn	{A^a_\nu}

\def \Abm	{A_{b\mu}}
\def \ABm	{A^b_\mu}

\def \Acn	{A_{c\nu}}

\def \Arond	{A^{\hspace{-1.5mm}^{^o}}}
\def \ABmrond	{{{\ABm}^{\hspace{-3.1mm}^{\vspace{2mm} o}}}}

\def \FaMN	{F_a^{\ \mu\nu}}

\def \FAmn	{F^a_{\ \mu\nu}}

\def \FAMN	{F^{a\mu\nu}}

\def \EAm	{E^a_\mu}
\def \EAmrond	{{{\EAm}^{\hspace{-3.1mm}^{^o}}}}
\def \EAn	{E^a_\nu}
\def \EAr	{E^a_\rho}

\def \EBn	{E^b_\nu}

\def \EDn	{E^d_\nu}
\def \EDr	{E^d_\rho}

\def \EcN	{E_c^\nu}
\def \gmn	{g_{\mu\nu}}

\def \grs	{g_{\rho\sigma}}

\def \gMN	{g^{\mu\nu}}

\def \gRS	{g^{\rho\sigma}}

\def \Gmn	{G_{\mu\nu}}
\def \Gmr	{G_{\mu\rho}}

\def \Grs	{G_{\rho\sigma}}

\def \Gsn	{G_{\sigma\nu}}

\def \GMnr	{\Gamma^\mu_{\ \nu\rho}}

\def \gMnr	{\gamma^\mu_{\ \nu\rho}}

\def \rmn	{r_{\mu\nu}}

\def \Rmn	{R_{\mu\nu}}

\begin{document}
\begin{titlepage}
\title{Einstein-\YM black hole solutions \\
in three dimensions.} 
\author{Vincent Brindejonc
\\ 
DAPNIA \\ 
Service de Physique Nucléaire\\ 
CEA Saclay 91191 Gif sur Yvette Cedex}  
\date{}
\maketitle 
\abstract{
In this paper, a procedure which gives Euclidean solutions of 3-dimensional Einstein-\YM 
equations when one has solutions of the Einstein equations is proposed. The method is
 based on reformulating \YM theory in such a way that it becomes a gravity. It is applied
 to find black hole solutions of the coupled Einstein \YM equations.}
\end{titlepage}
\vspace{2mm}
{\bf Introduction.}

Solutions of Einstein-\YM (EYM) equations have already been the motivation for several studies
 \cite{BAR88}. For most of them the ansatz used is a sophisticated version of the Reissner-Nordström 
solution. This solution teaches a lot about the properties of EYM black holes, but it requires a
 numerical evaluation in the final stage. Here, a method is proposed to build analytically a family
 of Euclidean solutions to these equations. In this report, the method is restricted to three
 dimensions and SU(2) is chosen as the \YM (YM) group.

The study of 3 dimensional gravity is important because it provides a compromise between the
 triviality of 2-d gravity and the intricacy of the 4-d one. Classically, the Riemann tensor
 is entirely determined by the Ricci tensor and by the scalar curvature, thus the solutions
 of the Einstein equations can be studied almost systematically \cite{DES84}. From the quantum
 point of view, the formalism proposed by Ach\`ucarro, Townsend, Witten  \cite{ACH86,WIT88}
 shows that 3-d quantum gravity is, in principle, integrable \cite{WIT88} albeit not trivial.
 It can serve as a good model for exploring the complexity of quantum gravity \cite{ASH96} or
 at least of quantum field theory in curved space-time.

The study of quantum properties by a functional integral means that one has to look for Euclidean
 solutions that will provide a starting point for a saddle point approximation. 
This method has shown its efficiency in the development of black hole thermodynamical properties
 and in quantum cosmology \cite{HAR83}.
An important step is to introduce non-trivial matter fields such as the YM field in this procedure.

In the present work, the key to finding Euclidean solutions for the EYM equations is based on
 describing the 3-d YM theory in terms of gauge invariant variables \cite{LUN92}. In this
 procedure, YM theory takes the form of a gravity. Thus, the EYM equations reduce to those
 of two coupled gravities. The similarity between the two sets of equations leads to a very
 simple ansatz which reduces the two coupled equations to one simple Einstein equation. 

We apply our ansatz to the Euclidean continuation of BTZ black hole which is a particularly
 interesting three dimensional solution \cite{BAN92} of the Einstein equations. The mass and
 the energy of the solution are calculated with the help of the quasi-local formalism developed
 in \cite{BRO93} and \cite{BRO94}.

\vspace{2mm}
{\bf Reformulation of the 3d SU(2) \YM theory as a Euclidean \\ gravity.}
\label{SECBASIC}

The reformulation of a YM theory as a strong gravity\footnote{``strong'' means that this gravity
 is not the universal one.} is based on the idea that there can exist a combination of YM fields
 which can be defined as a new metric on space-time. This is a very difficult task in four
 dimensions but some elegant ways exist in three dimensions such as, for example, the Lunev
 reformulation \cite{LUN92}. This reformulation is reviewed in the following, with a few modifications
 that will allow it to be adapted to our problem.
Consider the SU(2) YM theory written in first order formalism, on a 3-dimensional Euclidean manifold
 $\M$ with a metric $\gmn$:
\begin{eqnarray}
\label{EQSYMEU}
S_{\mathrm YM}=\frac{1}{16 \pi \as} \int_{\M}\d3x \rg  \Big( 2\FaMN (\partial_\mu \AAn -\partial_\nu
 \AAm-\epsilon^{abc} \Abm \Acn ) -\FaMN \FAmn \Big)
\end{eqnarray}
Where $\as$ is the square of the YM coupling constant.

Instead of $(\FAmn,\AAm)$ one can choose  $(\EAm , \AAm)$ as the fundamental variables where:
\begin{eqnarray}
\label{EQTRIAD}
\EAr=-\frac{1}{\as^2} {\rg \over 2} \epsilon_{\rho\mu\nu}\,\FAMN
\end{eqnarray}
The new variable $\EAr$ is simply the dual variable of $\FAmn$. The term $(\as)^{-2}$ which has, in 3d,
 the dimension of L$^2$, is added to the definition in order to obtain dimensionless quantities. This
 is necessary because we want to interpret 
\begin{eqnarray}
\label{EQMETFOR}
\Gmn=\EAm \,\delta_{ab} \, \EBn 
\end{eqnarray}
as a new metric field on space.  
The minus sign in Eq. (\ref{EQTRIAD}) is used in order to ensure that the strong gravitational constant 
will always have the same sign as the universal gravitational constant. In three dimensions there is, 
strictly speaking, no gravitational interaction and so a negative gravitational constant does  no 
harm\footnote{I thank G. Clément for pointing out that this question has already been settled in ref. 
\cite{DES84}.}.
If $\det(E)~>~0$ on $\M$ except eventually on a null measure set, then by substituting Eq. (\ref{EQTRIAD}) 
in Eq. (\ref{EQSYMEU}), $S_{\mathrm YM}$ can be rewritten\footnote{The reformulation is also possible when 
$\det(E)<0$, this gives a negative sign in front of $\sqrt{G}$ in Eq. (\ref{EQSGRA}). We have not yet 
explored this possibility.} as \cite{LUN92}
\begin{eqnarray}
\label{EQSGRA}
S_{\mathrm YM}=-\frac{\as}{4 \pi}\int_{\M}\d3x \Big\{ \rG R(\Gamma)+\frac{\as^2}{2}\rg \gMN \Gmn \Big\}
\end{eqnarray}
where $\Gmn$ is the strong metric defined in Eq. (\ref{EQMETFOR}), $\rG=\sqrt{\det(\Gmn)}$, $\Gamma$ is 
the strong gravitational connection defined by:
\begin{eqnarray}
\label{EQDEFGAM}
{\mathcal D}_\mu\,\EAn \equiv \partial_\mu\,\EAn - 
\epsilon^{abc} A_{b\mu} E_{c\nu}-\Gamma^\rho_{\nu\mu}E^a_\rho\equiv 0
\end{eqnarray}
and $R(\Gamma)$ is the scalar curvature of $\Gamma$.
One can see that the vacuum\footnote{Here, vacuum means absence of color sources.} YM equations can be 
written as: 
\begin{eqnarray}
\nonumber
\epsilon^{\lambda\mu\nu}D_\mu\,E^a_{\nu}=0
\end{eqnarray}
In conjunction with Eq. (\ref{EQDEFGAM}) this implies that if there are no YM sources\footnote{One can 
show that the universal gravity must also be torsion free.}, then $\Gamma$ is torsion free i.e.
\begin{eqnarray}
\label{EQTOR} 
T^\mu_{\,\nu\rho}(\Gamma)=0 
\end{eqnarray}
The first part of Eq. (\ref{EQSGRA}) is nothing but a pure Euclidean Einstein gravity action in its first 
order (or Palatini) formulation. The Euclidean character of this gravity is imposed by the signature of 
the su(2) Killing metric: $\frac{1}{2} \delta_{ab}$. One can remark that the Newton constant of this 
gravity is $\Gs = (4 \as)^{-1}$. The second part introduces a coupling between the new metric and the 
universal one. This term is not surprising because the YM action depends on the metric $\gmn$ which must 
appear somewhere in the reformulation. 

The reformulation can have several applications such as, for example, the study of the new degree of 
freedom $\Gmn$ and the study of confinement \cite{LUN96b}. We will concentrate on the search of 
solutions to the EYM equations where it appears to be very powerful.

\vspace{2mm}
{\bf Reformulated Einstein-\YM equations.}

We are now interested in the complete EYM action:
\begin{eqnarray}
\label{EQSEYM}
S= \int_{\mathcal M}  \d3x \rg \Big(\frac{1}{16\pi\as} \FAmn \FaMN - \frac{1}{16 \pi \Gn} (r(\gamma) +
 2 \Lam) \Big) + S_B
\end{eqnarray}
where $\Gn$ is the Newton constant, $r$ the scalar curvature of the connection $\gamma$, $\Lam$ the 
cosmological constant and $S_B$ the gravitational boundary term which will be detailed later on. If 
one uses the first order formalism for the YM part of the action one knows by Eq. (\ref{EQSGRA}) that:
\begin{eqnarray}
\label{EQSEYMR}
S= - \int_M  \d3x \Big(\frac{\rg}{16 \pi \Gn} (r + 2 \Lam)
+ \frac{\as}{4 \pi} \rG R(\Gamma) + \frac{\as^3}{8\pi}\rg \Gmn \gMN\Big) +S_B
\end{eqnarray}
Note that because of the sign in Eq. (\ref{EQTRIAD}) the strong gravity has the same sign as the 
universal gravity.\\
The equations of motion described in Eq. (\ref{EQSEYM}) are very asymmetric and thus lead to a 
very complicated ansatz \cite{BAR88}, however this is not the case for those described by Eq. (\ref{EQSEYMR}):  
\begin{eqnarray}
\label{EQCOMPg}
\rmn -\frac{1}{2} r \gmn -\Lam \gmn &=& -2 \Gn \as^3  (\Gmn -\frac{1}{2} \Grs \gRS \gmn) \\
\label{EQCOMPG}
\Rmn(\Gamma) -\frac{1}{2} R(\Gamma) \Gmn &=& \frac{\as^2}{2}  \sqrt{\frac{g}{G}} \Gmr \gRS \Gsn  \\
\label{EQCOMPgam}
\nabla^{\gamma}_\rho \gmn&=&0 \\
\label{EQCOMPGAM}
\nabla^{\Gamma}_\rho \Gmn&=&0 
\end{eqnarray}
Besides these equations of motion there are two conservation laws obtained by taking the covariant 
derivatives of Eqs. (\ref{EQCOMPg}) and (\ref{EQCOMPG}) with respect to $\gamma$ and $\Gamma$ 
respectively:
\begin{eqnarray}
\label{EQCONSg}
\nabla_{(\gamma)}^\mu\big(\Gmn -\frac{1}{2} \Grs \gRS \gmn\big)=0\\
\label{EQCONSG}
\nabla^{(\Gamma)}_\rho\big(\frac{\rg}{\rG}\gRS\big)=0
\end{eqnarray}
Equation (\ref{EQCONSg}) is the energy-momentum conservation law and it can be shown that Eq. 
(\ref{EQCONSG}) is the reformulated version of the YM Bianchi identities\footnote{The author thanks 
M. Knecht for this idea.}.

We will study the YM equations without a source term, then as seen in Eq. (\ref{EQTOR}), $\Gamma$ is 
torsion free and the complete set of Eqs. (\ref{EQCOMPg}-\ref{EQCOMPGAM}) reduces to:
\begin{eqnarray}
\label{EQREDg}
\rmn(g)&=& - 2 (\Lam \gmn + \Gn \as^3 \Gmn)\\
\label{EQREDG}
\Rmn(G)&=&\frac{\as^2}{2} \sqrt{{g\over G}}\gRS\big(\Gmr\Gsn-\Gmn\Grs\big)
\end{eqnarray}
\vspace{2mm}
{\bf A Simple ansatz.}

Since  Eqs. (\ref{EQREDg}) and (\ref{EQREDG}) are symmetric in $\Gmn$ and $\gmn$, the simplest 
ansatz consists by considering the two metrics as beeing related by a conformal transformation:
\begin{eqnarray}
\label{EQANSATZ}
\Gmn= \varphi^4(x) \gmn
\end{eqnarray}
By inserting Eq. (\ref{EQANSATZ}) in the Eqs. (\ref{EQCONSg},\ref{EQCONSG}) one immediately obtains
\begin{eqnarray}
\label{EQPHICON}
\partial_\mu \varphi = 0
\end{eqnarray}
which shows that $\varphi$ is constant. From Eq. (\ref{EQANSATZ}) and Eq. (\ref{EQPHICON}), it then 
follows that $\GMnr~=~\gMnr,\ \Rmn~=~\rmn$ and $R~=~r~/~\varphi^4$. Equations 
(\ref{EQREDg}) and (\ref{EQREDG}) becomes
\begin{eqnarray}
\label{EQTRANSFORMEg}
\rmn&=& - 2 \big( \Lam + \Gn \as^3 \varphi^4\big) \gmn\\
\label{EQTRANSFORMEG}
\varphi^2 \rmn&=& -\as^2\varphi^4\gmn
\end{eqnarray}
These equations show that $r$ is constant. Taking the trace of Eqs. (\ref{EQTRANSFORMEg},\ref{EQTRANSFORMEG}) one can see 
that,
\begin{itemize}
\item {if $r \geq 0$, the only possible solution for $\varphi$ is $0$. In this case, by Eq. 
(\ref{EQTRANSFORMEg}), $r=-6\Lam$ and then $\Lam\leq 0$. We then have a locally spherical space 
with no YM fields.}
\item {if $r \leq 0$, $\varphi$ has three possible values: $0$ and $\pm \sqrt{-r/(3\as^2)}$. The 
solution $\varphi=0$ corresponds to $\Lam \geq 0$ and then to a space which is locally the hyperbolic 
plane {\bf H}$^3$ and admits also BTZ black hole solutions. However, as in the previous case, this 
solution has no YM fields. The interesting solutions are $\varphi =\pm \sqrt{-r/(3\as^2)}$. They exist 
for $\Lam$ positive or negative and correspond to non-trivial persisting YM fields in the vacuum.}
\end{itemize}
\vspace {-2mm}
\begin {figure}[hhh]
\begin{center}
\mbox{\epsfig{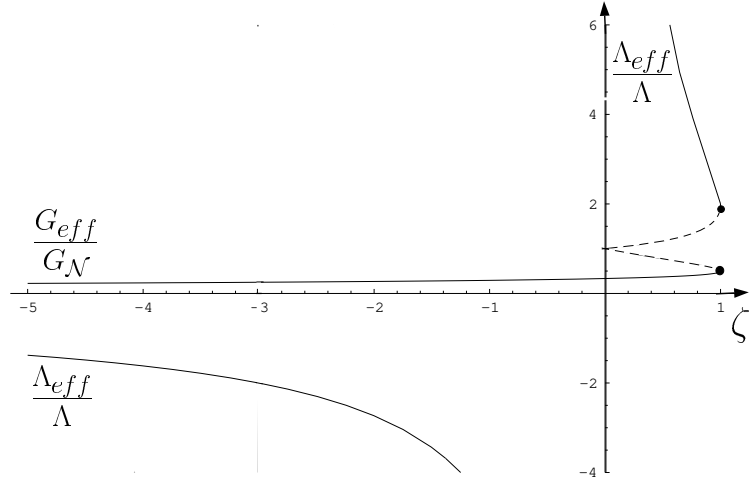}}
\end{center}
\caption{Plot of $\frac{\Lam_{eff}}{\Lam}$ and $\frac{G_{eff}}{\Gn}$ 
as a function of $\zeta$ for both cases ``$+$'' and ``$-$''. The solid and dashed line correspond 
respectively to ``$+$'' and ``$-$'' cases. The transitions between them takes place on the filled 
circles.}
\label{FIGEFF}
\end{figure}
The compatibility between Eqs. (\ref{EQTRANSFORMEg}) and (\ref{EQTRANSFORMEG}) implies that:
\begin{eqnarray}
\label{EQCOM}
\varphi^2&=&\frac{1}{4 \as \Gn} \Bigg( 1 \pm  \sqrt{1-\zeta}\ \Bigg)\\
\nonumber
\zeta&=& 16 \frac{\Gn \Lam}{\as}
\end{eqnarray}
In the following, we will use ``$+$'' to indicate the positive sign solution and ``$-$'' for the 
negative sign solution of Eq. (\ref{EQCOM}).
Since $\varphi^2$ must be positive if one wants to guarantee the positive sign of $\sqrt{G}$, the 
ranges of the dimensionless constant $\zeta$ are $]-\infty,1]$ for the ``$+$'' solution and $[0,1]$ 
for the ``$-$'' one\footnote{This remains true at least as long as $\Gn$ is positive.}. Then $\zeta$ 
can be viewed as the ratio of the interaction and vacuum characteristic lengths, i.e. $\zeta = 
\pm 16 l_\alpha l_{G_{\mathcal N}}/l_\Lam$.
When the compatibility condition (\ref{EQCOM}) is respected, there remains the Einstein equations 
to be solved:
\begin{eqnarray}
\label{EQEEFF}
\rmn=-2 \Lam_{eff} \gmn
\end{eqnarray}
An effective cosmological constant, induced by YM theory and gravity,
\begin{eqnarray}
\label{EQLAMEFF}
\Lam_{eff}= \Lam \frac{2 (1 \pm \sqrt{1-\zeta})}{\zeta}
\end{eqnarray}
has been introduced.
We have considered that $\Gn$ and $\as$ are positive quantities, therefore $\Lam_{eff}\geq 0$ for 
all values of $\Lam \leq \frac{\as}{16 \Gn}$.
Replacing $\Gmn$ by its ansatz in Eq. (\ref{EQSEYM}) one obtains in addition to $\Lam_{eff}$, an 
effective gravitational constant:
\begin{eqnarray}
\label{EQGEFF}
G_{eff}= \Gn \frac{1}{2 \pm \sqrt{1-\zeta}}
\end{eqnarray}
As can be seen in figure \ref{FIGEFF}, when $\zeta\in[0,1]$ the two solutions, ``$-$'' and ``$+$'' 
coexist. For these solutions $\Lam$ and $\Lam_{eff}$ are positive and the addition of the YM fields 
does not change the topology of space. In contrast to this, when $\zeta$ is negative, $\Lam_{eff}$ 
remains positive while $\Lam$ is negative, then the addition of the YM fields changes the topology 
of space from a quotient of S$^3$ to a quotient of {\bf H}$^3$.
One can note that when $\zeta\rightarrow 0$ the ``$-$'' solution gives 
\begin{eqnarray}
\nonumber
\Lam_{eff}\rightarrow \Lam \ \ , \ \ G_{eff}\rightarrow \Gn
\end{eqnarray}
This happens when one of the length scales defined by $\as$ or $\Gn$ tends to $0$ and therefore when 
the two theories decouple.

To summarize, one can say that the ansatz proposed in Eq. (\ref{EQANSATZ}) reduces the EYM equations 
to the Einstein equations with an effective cosmological constant (\ref{EQLAMEFF}) which in all cases 
is positive, corresponding to an attractive interaction, and an effective gravitational constant, 
(\ref{EQGEFF}).

When one has solved Eq. (\ref{EQEEFF}), one has a solution for $(\gmn,\Gmn)$ and consequently for 
$(\gmn,\EAm)$, where $\EAm$ is defined up to an SO(3) rotation. Using Eq. (\ref{EQDEFGAM}), one 
obtains the YM field $\AAm$ as being
\begin{eqnarray}
\label{EQDEFA}
\ABm={1\over 2} \epsilon^{bcd} \big( \EcN \partial_\mu \EDn - \EcN \Gamma^\rho_{\nu\mu} \EDr \big)
\end{eqnarray}
It is interesting to note that the solutions found through this reformulation are fully non-Abelian. 
This can be viewed as beeing due to the absence of a  Killing vector $\xi=\xi^a\sigma_a/2$ in the 
su(2) algebra. Such a Killing vector is defined by
\begin{eqnarray}
\nonumber
D_\mu \xi = 0
\end{eqnarray}
This is equivalent to $\epsilon_{abc}\EAm \xi^b=0$. This last relation shows that $\xi$ exists if, 
and only if, $\EAm$ is not invertible, i.e. if $\det(E)= 0$. The configurations given by this 
reformulation are thus totally non-Abelian, which is not the case of the currently known analytical 
solutions\footnote{A review of YM equations solutions can be found in \cite{ACT79}.}.

\vspace{2mm}
{\bf Example of an application: Black hole solutions.}

Many solutions to Eq. (\ref{EQEEFF}) are known but one of the most exciting ones is the BTZ
 black hole \cite{BAN92}. We will consider here the non-rotating BTZ black hole. One can extrapolate 
it to Euclidean time. It can be easily shown that through such a continuation, the Euclidean solution 
of the Einstein equations is:
\begin{eqnarray}
\label{EQMETBTZ}
ds^2 &=& \gmn dx^\mu dx^\nu \\
\nonumber
& =& N^2 d\tau^2 + \frac{dr^2}{N^2} + r^2 d\theta^2 
\end{eqnarray}
where $N$ is a function of the radial distance\footnote{In the following sections $r$ is only the 
radial distance and not the curvature of $\gamma$.} ``$r$'' such that
\begin{eqnarray}
\label{EQBTZN2NTH}
N^2(r)&=& - 8 G_{eff} M_{eff} + \Lam_{eff} r^2
\end{eqnarray}
Here, $M_{eff}$ is an integration constant.\\
The strong line element is 
\begin{eqnarray}
\label{EQSOLMETFOR}
dS^2 &=& \Gmn dx^\mu dx^\nu \\
\nonumber
& =& \varphi^4(N^2 d\tau^2 + \frac{dr^2}{N^2} + r^2 d\theta^2 )
\end{eqnarray}
and corresponds equally to the BTZ line element in the dilated space coordinate $dT=\varphi^2 d\tau , 
dR=\varphi^2 dr , d\Theta= d\theta$. As for the BTZ black hole solution, the line element, Eq. 
(\ref{EQMETBTZ}) has apparent singularities at
\begin{eqnarray}
\label{EQRSING}
r_s =\left(\frac{ 8 G_{eff} M_{eff}}{\Lam_{eff}}\right)^{1/2}
\end{eqnarray}
As can be seen in figure \ref{FIGEFF}, for fixed $\zeta>0$ and fixed positive $M_{eff}$ there are 
two black holes, a large one corresponding to the ``$-$'' solution in Eq. (\ref{EQCOM}) and a smaller 
one corresponding to the ``$+$'' solution. Both of these black holes have a radius smaller than the 
vacuum BTZ black hole. This can be interpreted by the fact that in this regime both the universal and 
strong gravities are attractive. \\
For fixed $\zeta<0$ there is only one black hole. In this case, strong gravity is attractive while 
universal gravity is repulsive but strong gravity wins and black holes can appear.

\vspace{2mm}
{\bf \YM fields.}

We will now show the YM fields corresponding to the strong metric, Eq. (\ref{EQSOLMETFOR}). This 
strong metric can be built from the particular triad field:
\begin{eqnarray}
\label{EQMATTRIAD}
\nonumber
\mu=\begin{array}{ccc}
0,&1,&2
\end{array} \hspace{4mm}&  \\
\EAmrond=\frac{2\Lam_{eff}}{\as^2}\left(
\begin{array}{ccc}
N & 0          & 0\\
0 &\frac{1}{N} & 0\\
0 & 0          & r
\end{array}
\right)
&
\hspace{-0cm}
\begin{array}{c}
a=1\\
a=2\\
a=3
\end{array}
\end{eqnarray}
However, when we decompose the strong metric, Eq. (\ref{EQMETFOR}) into a triad field we 
introduce a gauge choice in a gauge invariant formalism. The most general triad solution 
to our problem is
\begin{eqnarray}
\nonumber
\EAm = \Omega^{-1} \EAmrond {\hspace{3.1mm} \Omega}
\end{eqnarray}
where $\Omega$ is any SU(2)/${\mathbb Z}_2$ valued function.
Inserting Eq. (\ref{EQMATTRIAD}) into Eq. (\ref{EQDEFA}) and keeping the same conventions for 
the matrices, one obtains:
\begin{eqnarray}
\nonumber
\ABmrond \hspace{3.1mm} =\left(
\begin{array}{ccc}
0     & 0 & -N \\
0     & 0 & 0 \\ 
N  N' & 0 & 0
\end{array}
\right)
\end{eqnarray}
where $N'$ represents the derivation of $N$ with respect to $r$. Taking Eq. (\ref{EQBTZN2NTH}) 
into account this reduces to
\begin{eqnarray}
\nonumber
\ABmrond \hspace{3.1mm}=\left(
\begin{array}{ccc}
0            & 0 &-N \\
0            & 0 & 0 \\
\Lam_{eff} r & 0 & 0
\end{array}
\right)
\end{eqnarray}
This result can be rewritten in the gauge algebra as:
\begin{eqnarray}
\label{EQBBMU}
\Arond \hspace{1mm} \equiv &\ABmrond \hspace{3.1mm} \frac{\sigma_b}{2}dx^\mu
\end{eqnarray}
\begin{eqnarray}
\nonumber
=\left(
\begin{array}{cc}
r \Lam_{eff} d\tau & - N d\theta\\
 & \\
- N d\theta &  - r \Lam_{eff} d\tau
\end{array}
\right)
\end{eqnarray}
where $\sigma_b$ are the Pauli matrices:
\begin{eqnarray}
\nonumber
\sigma_1=\left(
\begin{array}{cc}
0 & 1 \\
1 & 0  
\end{array}
\right)
&
\sigma_2=\left(
\begin{array}{cc}
0 & - i  \\
i & 0  
\end{array}
\right)
&
\sigma_3=\left(
\begin{array}{cc}
1 & 0 \\
0 & -1  
\end{array}
\right)
\end{eqnarray}
In the same way as for $\EAmrond\ $, the solution $\Arond$ can be extended to
\begin{eqnarray}
\nonumber
A=\Omega^{-1}\Arond \hspace{3mm} \Omega+i\Omega^{-1}\partial_\mu \Omega
\end{eqnarray}
Note that all the possible solutions of $A$ are classified by the homotopy classes of the 
applications $\Omega:{\mathcal M}\rightarrow {\mathrm SO(3)}$. In thermodynamical applications, 
time is ``compactified'' and the interior region has to be removed in such a way that the topology 
of ${\mathcal M}$ reduces to the empty torus. Finally, the possible solutions for the potential, $A$, 
are classified by the homotopy classes of $\Omega:S^1\times S^1 \rightarrow {\mathrm SO(3)}$.

\vspace{2mm}
{\bf Energy and mass.}

In this section the energy and the mass of the black holes found before are given. We will use the 
quasi-local formalism of \cite{BRO93}, specially extended in \cite{BRO94} to the case of the BTZ 
black hole. This formalism is powerful because it can be applied in a bounded, finite spatial region, 
with no assumptions on the asymptotic flatness of space-time. It is based on a Hamilton-Jacobi 
treatment of the gravitational action, Eq. (\ref{EQSEYM}) with its boundary term 
\begin{eqnarray}
\nonumber
S_B=\frac{1}{8 \pi \Gn} \int_{\Sigma_{t'}\cup\Sigma_{t''}}{d^2x \sqrt{h} K}-\frac{1}{8 \pi \Gn} 
\int_{\mathcal{B}}{d^2x \sqrt{-\gamma} \Theta}
\end{eqnarray}
In this equation $\Sigma_t$ is a space-like hyper-surface defined by $t=$constant and $r\leq R$, 
$\mathcal B$ is the surface defined by $t' \leq t \leq t''$ and $r=R$. $h_{ij}$ and $\gamma_{ij}$ 
are respectively the metrics on $\Sigma$ and $\mathcal B$ with external curvatures $K$ and $\Theta$. 
For a given solution one varies this action but leaves the metric on the boundaries free. In this 
case one obtains 
\begin{eqnarray}
\nonumber
\delta S= \delta S_{\mathrm YM} + \int_{\Sigma_{t'}\cup\Sigma_{t''}}{d^2x P^{ij} \delta h_{ij}}+ 
\int_{\mathcal{B}}{d^2x \pi^{ij} \delta \gamma_{ij}}
\end{eqnarray}
where $P_{ij}$ is the gravitational momentum associated with the space-like hyper-surface $\Sigma$ 
and $\pi_{ij}$ the gravitational momentum on $\mathcal B$. This last momentum can be interpreted as
 a quasi-conserved energy-momentum tensor on $\mathcal B$. Selecting its normal-normal, normal-parallel 
and parallel-parallel components one can extract the quasi-local energy $\epsilon$, the angular
 momentum $J_\theta$ and the stress $s^{\theta\theta}$. All these quantities are defined up to an 
additive constant because one can always add on a boundary action $S^0$ to $S$. This procedure 
fixes the zero energy point. Finally, if we apply this formalism to the Minkowskian continuation of our 
metric, one has \cite{BRO94}
\begin{eqnarray}
\label{EQENERGY}
\epsilon&=&\frac{\sigma^{\theta\theta}k_{\theta\theta}}{8 \pi \Gn}-\epsilon_0\\
\label{EQANGMOM}
\j_{\theta}&=&0\\
\label{EQSHEAR}
\sigma_{\theta\theta}s^{\theta\theta}&=&-\frac{1}{8 \pi\Gn} n_\mu a^\mu-
\frac{\partial R \epsilon_0}{\partial R}
\end{eqnarray}
where $\sigma_{\theta\theta}= R^2$ is the metric\footnote{One has to consider here the Minkowskian
 metric. We use the signature convention (-,+,+).} on $B=\partial \Sigma$, $k_{\theta\theta}=-R N$ 
is the extrinsic curvature of $B$, $\hat n$ is the normal vector of $\mathcal B$ and $a^\mu$ is the
 acceleration of $B$. The choice of the subtraction term $\epsilon_0$ poses a problem. In the range 
$\zeta\in[0,1]$ it is possible to take the empty anti-De Sitter (aDS) space-time as a reference but 
it is no longer possible in the range $\zeta\in]-\infty,0[$ where the empty space-time is De Sitter 
and where the energy is not defined beyond the cosmological horizon\footnote{It is of course always 
possible to take a flat space-time ($\epsilon_0=0$) as a reference, but this is not physically well 
defined.}.

In the case where $\Lam>0$, if one takes as a reference the space-time metric of the empty aDS 
space-time, then the total internal energy is 
\begin{eqnarray}
\label{EQENERGYTOT}
E&=&\int_B {d\theta \sqrt{\sigma} \epsilon}\\
\nonumber
 &=&-\frac{1}{4 \Gn}\Big(\sqrt{-8 M_{eff} G_{eff} + \Lam_{eff} R^2}-\sqrt{1+\Lam R^2}\Big)\ ,
\end{eqnarray}
which diverges when $R\rightarrow\infty$, we then have
\begin{eqnarray}
\nonumber
E\sim -\frac{1}{4 \Gn}\Big(\sqrt{\Lam_{eff}}- \sqrt{\Lam}\Big) R
\end{eqnarray}
This infinity shows that our solution considerably changes the structure of space-time. Alternatively, 
if one wants a correctly referenced space-time for all values of $\zeta$ one has to take the aDS 
space-time defined by $\Lam_{eff}$. Then for all $\zeta$,
\begin{eqnarray}
\nonumber
E&=&-\frac{1}{4 \Gn}\Big(\sqrt{-8 M_{eff} G_{eff} + \Lam_{eff} R^2}-\sqrt{1+\Lam_{eff} R^2}\Big)\ ,
\end{eqnarray}
which tends to zero as $R$ tends to infinity. 
The stress $\sigma_{\theta\theta}s^{\theta\theta}$ is proportional to the thermodynamical ``surface'' 
pressure on the 1-dimensional ``surface'' $B$:
\begin{eqnarray}
\nonumber
\sigma_{\theta\theta}s^{\theta\theta}=-\frac{1}{8 \pi \Gn}(N'(R)-N_0'(R))
\end{eqnarray}
where the $N'$ denotes the derivation of $N$ with respect to $R$ and where $N_0$ is the subtraction term 
$\sqrt{1+\Lam_0 R^2}$ with $\Lam_0=\Lam$ or $\Lam_{eff}$ depending on the background choice. This 
``surface'' pressure goes as 
\begin{eqnarray}
\nonumber
\sigma_{\theta\theta}s^{\theta\theta} \sim_{\hspace{-5mm} _{R\rightarrow \infty}} & -\frac{1}{8 \pi \Gn} 
(\sqrt{\Lam_{eff}}-\sqrt{\Lam_{0}})
\end{eqnarray}

In the quasi-local formalism, the conserved mass is not equivalent to energy\footnote{Energy is not a 
conserved quantity here \cite{BRO93}.} in an asymptotically non-flat space-time. It is associated with 
the conserved charge of the time-like Killing vector and is thus (\cite{BRO94})
\begin{eqnarray}
\label{EQMASS}
M&=&\int_B {d\theta \sqrt{\sigma} N \epsilon}\\
\nonumber
 &=&-\frac{N(N-N_0)}{4 \Gn}
\end{eqnarray}
and with our choice, $\Lam_0=\Lam_{eff}$, this goes as\footnote{The difference with the BTZ
 mass given in \cite{BAN92}, comes from our choice for the background space of an aDS space 
and not of the space obtained in the limit $M_{eff}=0$.}
\begin{eqnarray}
\nonumber
M & \sim_{\hspace{-5mm} _{R\rightarrow \infty}} &\frac{M_{eff}G_{eff}-1}{\Gn}
\end{eqnarray}
This last value gives the pure gravitational mass of the black hole.

\vspace{2mm}
{\bf Conclusions and outlook.}

In the previous sections we have developed a new method for finding solutions to the 3-d Euclidean 
EYM equations. The crux of this method is the use of a  reformulated YM theory as a gravitational 
theory. The present reformulation suggests many ansätze. We have studied the simplest one based on 
a conformal link between the strong and the true metric. This ansatz is exactly solvable for the YM 
field when one has a solution to the Einstein equations. 

The ansatz gives some positive results: the solutions of EYM are explicit, they are fully non-Abelian, 
they reveal a rich topological structure of EYM space\footnote{These two points are true in general 
in the reformulation framework.} due to a non-trivial mapping between the base space and the gauge 
group. It is possible that this topological structure is linked with the classification of the usual 
EYM black holes by a set of integer numbers \cite{BAR88}.

Unfortunately, this ansatz is very rough and it leads to a bad asymptotic behaviour for the YM field. 
In the case of the black hole solutions, we have found that the bad behaviour of the YM fields greatly 
changes the topology of space. For this reason, it is not possible in the case where $\Lam < 0$ to
 choose the vacuum De Sitter space as a reference space. When $\Lam \geq 0$, the topology of space 
remains unchanged and it is, in principle, possible to take an aDS space as a reference. However, 
due to the shift from $\Lam$ to $\Lam_{eff}$, all the quantities obtained in this case are infinite. 
This point is not yet physically understood. Nevertheless, it is possible to begin a study of the 
thermodynamical quantities of these black holes using Eq. (\ref{EQENERGYTOT}) and those thereafter. 
The charge of the black hole is still unclear \cite{CLE97} and so we have not included it in this 
discussion. 

The bad behaviour of the YM fields can be cured in extended ansätze. The conformal ansatz considered 
here belongs to the large ensemble where the two metrics are related by a general transformation $A$:
\begin{eqnarray}
\nonumber
\Gmn(x)=A^\rho_\mu(x)\grs(x)A^\sigma_\nu(x)
\end{eqnarray}
One could constrain $A$ by requiring a good asymptotic behaviour.

Many other extensions are possible. We have applied our ansatz to a BTZ black hole, but this can be 
easily extended to all vacuum solutions of the Euclidean Einstein equations in three dimensions such 
as multi-black hole solutions \cite{CLE94}. Moreover, as our procedure systematically eliminates the 
gauge connection of the equations of motion, one can include various other fields such has dilaton,
 electro-magnetic, etc. The inclusion of other fields directly coupled to gauge fields, such as quark 
fields, is more difficult but possible in the framework of the ansatz. At the moment, this reformulation 
is well defined only in three dimensions with a gauge group SU(2). It is however possible to easily
 include matter sources which will give torsion and a topological term to the YM theory which will in 
turn give a strong gravitational topological term. It is important to be assured that the reformulation 
is not an accident of SU(2) gauge theories in three dimensions. It is in fact possible to extend this 
reformulation to the SU(N) gauge group in three dimensions but the result is already rather complicated 
for SU(3). There are also some possibilities for reformulating in 2 and 4 dimensions \cite{LUN96a}.

\vskip 0.8cm 
\noindent{\bf Acknowledgments} \\
\noindent I want to thank in particular M. Bauer for the stimulating discussions we had and all his 
explanations necessary to this article. It is a pleasure to thank G. Cohen-Tannoudji for his advice 
and critical reading, M. Knecht, G. Clement and S.deser for their remarks, M. Mac Cormick for her 
corrections and I. Porto-Cavalcante and D. Lhuillier for their stimulating discussions.

{} 
\end{document}